\documentclass{article}
\usepackage{mathtools,xspace}
\usepackage{tikz}

\title{Answering Conjunctive Queries over Linear Existential Rules with Transitivity}

\author{C. Feier and T. Gogacz and F. Murlak} 

\begin{document}

\maketitle

\begin{abstract}
    We study Conjunctive Query Entailment Problem over  ontologies consisting of Linear Tuple Generating Dependencies and transitivity assertions. A special case of this problem has been shown to be decidable in \cite{baget2015combining}. In particular, it covers the case of linear single head TGDs with binary signature. In This paper we show that The problem for linear multi head TGDs with binary signature is Ackermanian hard
\end{abstract}
\section{Draft of the lowerbound}
We know by \cite{lossy} \footnote{for a detailed description one can look to: mimuw.edu.pl/$\sim$sl/teaching/21\_22/JAiO-II/lossy.pdf} 
that reachability for lossy counter machines is Ackermann-hard. Transitions of a counter machine, besides changing the control state, perform a specified operation on a counter:
increment by one, decrement by one, or zero test. The word 'lossy' means, that counters can spontaneously decrease at any time.

We want to reduce this problem to query entailment for LinTrans ontologies, although we will not do this in full generality. We will reduce only those instances that are in the image $B$ of the reduction in \cite{lossy}. This suffices to show hardness. Moreover we can exploit additional properties of automata in $B$: all accepting runs are loss free (i.e. no spontaneous counter decreese). Due to this property our reduction does not have to implement all possible spontaneous counter decreases. In fact, it could have no possible spontaneous decreases at all.  But we can allow for arbitrary decreases to happen as artifacts of the construction. If there is an accepting run, then it is loss-free so we will find it. If there is no accepting run, then we will not get one due to lossyness existing in our reduction, since even full lossyness does not introduce false accepting runs. 

Let $\mathcal{Q}$ be a set of control states of lossy $k$-counter machine. Let $q_I,q_F$ be initial and final states, respectively. We ask if there is a run from the initial configuration $(q_I,\langle 1,0,0,\ldots,0\rangle )$ to the final configuration $(q_F,\langle 1,0,0,\ldots,0\rangle )$.

We will test the entailment by the query rewriting method. Therefore one should think that linear rules are {\bf applied backwards}. The linear rules will be written in a standard way, but semantics coincides with backward reading.

A configuration of a lossy $k$-counter machine will be represented by the following structure. A central node (where computation happens) and $k+1$ paths connected to the center. $k$ paths will encode counters, $1$ path will keep track of the state.

We would like to represent a counter with value $l$ by path of length $l$. Unfortunately, such a simple representation, together with the rest of our construction, would allow for spontaneous counter increases. 
Not just spontaneous counter increases. Therefore for each counter we will have two atoms that encodes the parity of the token on the counter. We will interpret the number of alternating blocks of different parity as the value of a counter. It is easy to see, that the number of alternating parity blocks does not change with local duplication of an edge. 
For the technical reason, we introduce a third atom for each of the counters that denotes the end of the counter stack.

For the state path we will have (ideally) single atom encoding the (annotated) state , followed the end of path atom.

For each counter $i\in \{1,2,\dots,k\}$ we introduce three transitive binary predicates $1^i,0^i,e^i$. They represent odd tokens, even tokens, and the bottom token for counter $i$. For each state $q$ we introduce predicates $q^{B_1,B_2,\ldots, B_k}$ and 
$^iq^{B_1,B_2,\ldots, B_k}$, as well as a single predicate $e^{\mathrm{st}}$ that plays a role similar to $e^i$, but for the state, rather than register $x_i$.  The right superscript indicates the parity of tokens with which the machine should interact next in each counter (the bottom symbol is considered even). The left superscript $i$ in the second predicate indicates the counter to be increased as second substep of a transitions.

\begin{figure}
\label{fig:queries}
\begin{tikzpicture}
\newcommand{\exampl}[3]{

\node[draw,circle](C) at (0+#1,2+#2) {};
\node[draw,circle](l1) at (0+#1,-1+#2) {};
\node[draw,circle](l2) at (1+#1,-1+#2) {};
\node[draw,circle](l3) at (2+#1,-1+#2) {};
\node[draw,circle](l4) at (3+#1,-1+#2) {};
\node[draw,circle](l5) at (4+#1,-1+#2) {};
\node[draw,circle](l1e) at (0+#1,-2+#2) {};
\draw (C) -- (l1)  node[left, pos=0.8] {$1^1$};
\draw (C) -- (l2)  node[left, pos=0.8] {$e^2$};
\draw (C) -- (l3)  node[left, pos=0.8] {$e^3$};
\draw (C) -- (l4)  node[left, pos=0.8] {$e^4$};
\draw (C) -- (l5)  node[left, pos=0.8] {$e^5$};
\draw (l1) -- (l1e)  node[left, pos=0.5] {$e^1$};

\node[draw,circle] (Q) at (2+#1,3+#2) {};
\node[draw,circle] (Qe) at (3+#1,3.5+#2) {};

\draw (C) -- (Q)  node[left, pos=0.8] {$q_#3^{1,0,0,0,0}$};
\draw (Q)-- (Qe) node[left, pos=0.8] {$e^{\mathrm{st}}$} ;
}
\exampl{0}{0}{I};
\exampl{6}{0}{F};

\end{tikzpicture}
\caption{ $Q_I$ and $Q_F$}
\end{figure}
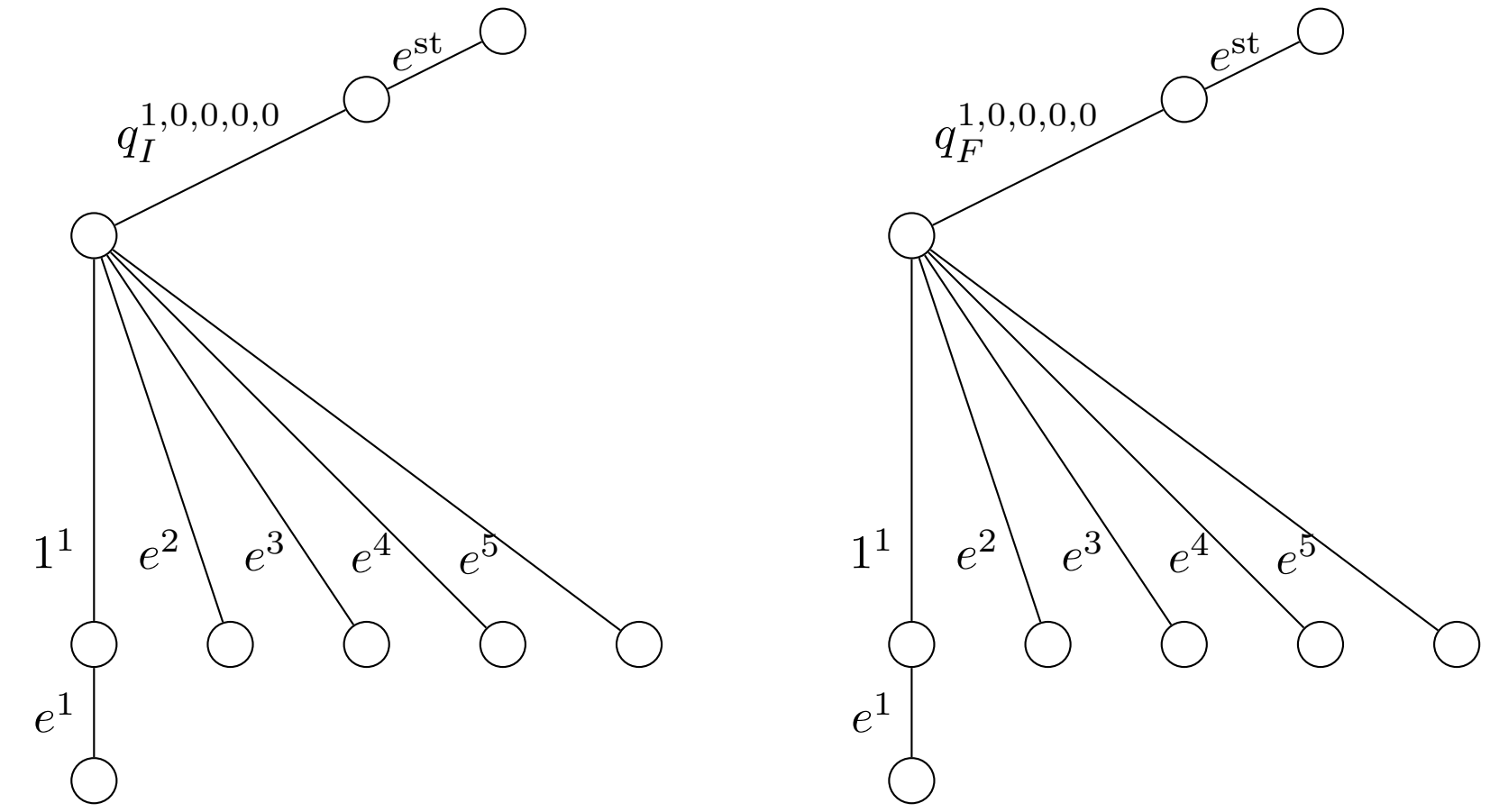

Figure \ref{fig:queries} shows the encoding of the initial and the final configuration of the encoded lossy automaton.

We will prove that there is a run from the initial configuration to the final one iff the query $Q_I$  can be backward rewritten into the query $Q_F$. In the standard terms: $Q_F,{\cal T}\models Q_I$.

The implementation of a machine transition will depend on the type of counter manipulation associated with it. We write $^*\!q^{B_1, B_2, \dots, B_k}$  for a restricted wildcard that can be replaced with $q^{B_1, B_2, \dots, B_k}$ or $^{i'}\!q^{B_1, B_2, \dots, B_k}$ for an arbitrary $i'$.

Initially, for $q \xrightarrow{x_i=0} p$, $q \xrightarrow{x_i++} p$, $q \xrightarrow{x_i--} p$  we get respectively:

\begin{multline}p^{B_1,B_2,\ldots, B_k} (x,y)\longrightarrow \exists z,u \;\; e^{\mathrm{st}}(u,y) \wedge\;^*\!q^{B_1,B_2,\ldots, B_k}(z,u)\wedge \bigwedge_{j\neq i} (B_j)^j(x,z)\wedge e^i(x,z) \label{counter_test}
\end{multline}

\begin{multline}^ip^{A_1,A_2,\ldots, A_k} (x,y)\longrightarrow \exists z,u \;\; e^{\mathrm{st}}(u,y) \wedge\;^*\!q^{B_1,B_2,\ldots, B_k}(z,u)\wedge \bigwedge_{j} (B_j)^j(x,z)\\ \text{where } B_i:=1-A_i  \text{ and }   B_j:=A_j \text{ for }j \in \{1, 2, \dots, k\} \setminus \{ i \}, \label{counter_increase}
\end{multline}

\begin{multline}p^{A_1,A_2,\ldots, A_k} (x,y)\longrightarrow \exists z,u \;\; e^{\mathrm{st}}(u,y) \wedge\;^*\!q^{B_1,B_2,\ldots, B_k}(z,u)\wedge \bigwedge_{j} (B_j)^j(x,z)\\ \text{where } B_i:=1-A_i  \text{ and }   B_j:=A_j \text{ for }j \in \{1, 2, \dots, k\} \setminus \{ i \} \label{counter_decrease}
\end{multline}

To get the full set of linear rules of the types (1)-(3), one have to do all possible substitutions: the $e$ predicate can be used as generic even predicate in place of $0$.

Additionally, to facilitate counter incrementation required after rule \ref{counter_increase} we introduce the following linear rules for all $i\in\{1, 2, \dots, k\}$ and all $B_1, B_2, \dots, B_k \in \{0,1\}$,
\begin{equation}
(B_i)^i(x,z) \longrightarrow \exists y\;\; ^ip^{B_1,B_2,\ldots, B_k}(y,z) \wedge (1-B_i)^i(x,y)  \land \bigwedge_{j\neq i} (B_j)^j(z,y)\,. \label{counter_increase_implementation}
\end{equation}
Although at first glance these rules `consume' tokens, in the intended semantics these tokens will be duplicated beforehand  by a backward application of the transitive rule.

\textcolor{red}{The rule is intended to fire exactly once. If it fires more than once (because the token on the left was first duplicated many times), it will break the alternation of 0 and 1. (When is it actually checked?) If it fires zero times, this corresponds to the toked being lost, which is allowed in a lossy counter machine. This means that the tokes are not really lost, but rather not put on the stack at all.}

First, assume that there is an accepting run. We will show how to rewrite the initial query into the final one. This description also serves as a demonstration of the intended use of rules \eqref{counter_test}--\eqref{counter_increase_implementation}. Notice that the lossy counter machine we are implementing has accepting run only if there was not a single loss of token. Therefore we need to check that every non-lossy step can be simulated.

Notice that rules  (\ref{counter_test}--\ref{counter_increase_implementation}) always consume tokens from all counters. This behavior is unwanted. So before we apply any of those rules, we split token atoms of counters that should not be decreased. 
if we used rule (\ref{counter_increase}), next we split the state atom. Afterwards use rule (\ref{counter_increase_implementation}) to increase appropriate counter.
By induction we prove that the counters content is correct so we will obtain the final query.

Second, assume that we can rewrite the initial query into the final one. We want to show that there is an accepting run of the lossy counter machine. We will say that the $i$-th counter has value $n$ if there are $n$ alternations between $0$ and $1$ on the $i$-th path. We want to show that after any rewriting step we will get a query corresponding to a reachable step of the lossy counter machine. It is clear that the intended usage of rules is ok. We need to investigate the unintended cases.

Case 1: Spontaneous split of token atom. With our semantics this does not increase the value of counters, since we did not increase the number of alternations. 

Case 2: We did not split a token atom that was not supposed to be erased. It is ok, since this corresponds to a spontaneous decrementation of a counter, and our machine is lossy.

Case 3: We did split a token that was supposed to be erased (not necessarily just before decrease step). Then parity of atoms on this path stays the same. But the necessary parity in the state atom changes.  We cannot therefore apply any further rule that changes the state, so we would not be able to get the final query.

Case 4: The state atom is split too many times, so the rule $\ref{counter_increase_implementation}$ is applied many times. This is ok, because the tokens added have the parity. So it is just counter+1 in our semantics.

\bibliographystyle{abbrv}
\bibliography{lower}{}

\end{document}